\documentclass[11pt,a4paper]{article}
\usepackage{amsmath,amsthm,latexsym,amssymb,amsfonts,parskip}
\usepackage{palatino}
\usepackage[mathcal]{euler}

\newtheorem{lemm}{Lemma}[section]
\newtheorem{prop}[lemm]{Proposition}
\newtheorem{coro}[lemm]{Corollary}

\newtheorem{assu}[lemm]{Assumption}

\newcommand{\scpr}[2]{\langle#1\, \vert \, #2 \rangle}
\newcommand{\scprd}[2]{(#1\, \vert \, #2 )}

\newcommand{\betr}[1]{\left\lvert #1 \right\rvert}
\newcommand{\poi}[2]{\left\{#1\,,\, #2 \right\}}
\newcommand{\R}{\mathbb{R}}                  % the reals
\newcommand{\Z}{\mathbb{Z}}                  % the integers
\newcommand{\N}{\mathbb{N}}                  % the natural numbers
\newcommand{\uone}{\text{U(1)}}
\newcommand{\RB}{\overline{\R}_{\text{B}}}
\DeclareMathOperator{\cyl}{Cyl}              % Cyl
\DeclareMathOperator{\diff}{\text{Diff}}
\DeclareMathOperator{\gs}{\text{GS}}
\DeclareMathOperator{\tdiff}{\text{TDiff}}
\newcommand{\hkin}{\mathcal{H}_{\text{kin}}}
\newcommand{\hdiff}{\mathcal{H}_{\text{diff}}}
\newcommand{\ket}[1]{\lvert #1 \rangle}

\newcommand{\brad}[1]{( #1 \rvert}
\newcommand{\ketd}[1]{\lvert #1 )}
\newcommand{\avec}[1]{{#1}_\cdot}

\newcommand{\ann}[1]{a_{#1}}
\newcommand{\cre}[1]{a^\dagger_{#1}}
\newcommand{\ind}{\mathcal{I}}
\newcommand{\kron}[2]{\delta(#1,#2)}
\newcommand{\labe}[1]{{#1}_\cdot}
\newcommand{\lab}[2]{{#1}_{#2}}
\newcommand{\dual}[1]{\widetilde{#1}}
\newcommand{\comm}[2]{[#1,#2]}
\newcommand{\wh}[1]{\widehat{#1}}

\title{Exploring the diffeomorphism invariant Hilbert space of a scalar field}
\author{Hanno Sahlmann, Institute for Theoretical Physics, Utrecht University}
\date{{\small Preprint ITP-UU-06/35, SPIN-06/31}}

\begin{document}
\maketitle
\begin{abstract}
As a toy model for the implementation of the diffeomorphism
constraint, the interpretation of the resulting states, and the
treatment of ordering ambiguities in loop quantum gravity, we
consider the Hilbert space of spatially diffeomorphism invariant
states for a scalar field. We give a very explicit formula for the
scalar product on this space, and discuss its structure.\newline
Then we turn to the quantization of a certain class of
diffeomorphism invariant quantities on that space, and discuss in
detail the ordering issues involved. On a technical level these
issues bear some similarity to those encountered in full loop
quantum gravity.
\end{abstract}
%------------------------------------------------------------------
\section{Introduction}
%------------------------------------------------------------------
The space of spatially diffeomorphism invariant states, $\hdiff$, is
important in loop quantum gravity (LQG): It may be home to the physical states of the
theory, and it is the space on which the Hamiltonian constraint,
arguably the most important operator of the theory, is defined. We
feel however that $\hdiff$ is not very well understood. For example
elements of $\hdiff$ are obtained by a group averaging procedure that
is quite subtle \cite{Ashtekar:1995zh,Ashtekar:2004eh}. Very roughly speaking,
they can be labelled by, among other things, diffeomorphism equivalence
classes of graphs, i.e. objects that are hard to describe \textit{explicitly}.
Also the physical meaning of the states is rather unclear.
One would need operators to ``probe" the states, operators that demonstrably
correspond to classical quantities of interest. But besides the total
spatial volume and the Hamilton constraint with constant lapse, there are
no candidates for such operators.

Thiemann \cite{Thiemann:1996aw,Thiemann:1997rt} \textit{has} given a
prescription of how to quantize a large class of quantities on
$\hdiff$, but there are issues (ex. \cite{Lewandowski:1997ba}), and
ambiguities whose physical meaning, and mathematical consequences
are not clear. There are very interesting and encouraging results on
some of them \cite{Bojowald:2002ny,Perez:2005fn}; see also
\cite{Thiemann:2006cf} for a concise discussion of the ambiguities.
But given the importance, specifically, of the \textit{correct}
implementation of the Hamiltonian constraint, we think it fair to
say that one knows too little.

Here we would like to start consideration of a toy model in which
the above-mentioned points can be studied with relative ease. More
precisely, we will study a scalar field in the \textit{polymer
representation}
\cite{Thiemann:1997rq,Starodubtsev:2002xe,Ashtekar:2002vh}. We will
not discuss the dynamics of the field. See \cite{Han:2006iq} for a
treatment of a \textit{physical} scalar field coupled to gravity.
The basic field quantities derived from the canonical pair
$(\phi,\pi)$ that are subject to quantization are
\begin{equation}
\label{eq_basic} T_{x,\alpha}=\exp [i\alpha \phi(x)],\qquad
\pi(f)=\int \pi(y) f(y),\qquad \alpha\in \ind.
\end{equation}
We will consider two cases, $\ind=\Z,\R$.\footnote{As will become apparent
when we introduce the representation for these quantities, it is mathematically
more appropriate to describe $\ind$ as the Pontryagin dual of U(1),
and of $\RB$ (the Bohr compactification of $\R$), respectively.}
In the former it may be
more natural to consider the $T_{x,\alpha}$ as functions of the U(1)
valued field $T_{x,1}$, forgetting about $\phi$
altogether.\footnote{After all, in this case the $T_{x,\alpha}$ do
not suffice to reconstruct $\phi$.}

In the first part of the paper we discuss the construction of the
diffeomorphism invariant Hilbert space for the field in analogy to
that of LQG. Making a certain assumption on the class of allowed
diffeomorphisms enables us to characterize states and scalar product
very explicitly. It turns out that the Hilbert space can be
characterized as a Fock-space in a natural way.

In the second part of the paper, we will be concerned with the
quantization of certain diffeomorphism invariant quantities,
\begin{equation}
\label{eq_thel} L_\alpha\doteq\int \pi(x) \exp[ i\alpha \phi(x)] \qquad
\alpha\in\ind.
\end{equation}
One reason to consider these quantities is that they are easily
expressed in terms of those in \eqref{eq_basic}, and thus
quantization may be expected to be relatively straightforward.

Another reason is that the ordering problems that can be expected for
the quantization of \eqref{eq_thel} are analogous to (though much
simpler than) those encountered for the ``$FEE$" term in the
Hamiltonian constraint.

Finally, an important reason is that they form an algebra under
Poisson brackets:
\begin{equation}
\label{eq_witt_poi}
\poi{L_\alpha}{L_{\alpha'}}=i(\alpha-\alpha')L_{\alpha+\alpha'}.
\end{equation}
For $\ind=\Z$ one recognizes the Witt algebra, the algebra of
vector-fields on the circle. Direct calculation confirms that in this
case the $L_\alpha$ generate diffeomorphisms of the target $\uone$.
For $\ind=\R$ one may therefor think of the $L_\alpha$ as generating
``diffeomorphisms" of $\RB$, but we will not bother to give this a
technical meaning.

Thus for the quantities \eqref{eq_thel} there is a
simple and thorough test of the quantization: Is \eqref{eq_witt_poi}
reproduced (in appropriate commutation- and adjointness relations)?

We will unfortunately not be able to find a complete solution to this
quantization problem in the present paper. That certainly does not
mean that there is none. The structures and obstruction we find will
however show that this problem is subtle and merits further
investigation.

The paper is organized as follows:\\
Section \ref{se_hkin} recalls the kinematical quantization of a
scalar field in LQG.\\
Section \ref{se_hdiff} gives an explicit description of the
Hilbert-space of diffeomorphism invariant states $\hdiff$ for the
scalar field.\\
In Section \ref{se_quant}, we will be concerned with the quantization
of the $L_\alpha$.\\
We end the paper with a discussion of results and possible future
work in Section \ref{se_close}.\\
An appendix contains several longer computations.

%Conventions: We use conventions such that
%$\poi{\phi(x)}{\pi(y)}=\delta(x,y)$, and
%$\comm{\widehat{\phi}(x)}{\widehat{\pi}(y)}=i \delta(x,y)$, the
%latter in a heuristic sense, since $\widehat{\phi}$ will not exist.

%------------------------------------------------------------------
\section{Kinematical quantization}
\label{se_hkin}
%------------------------------------------------------------------
In this section we recall the standard quantization of a scalar
field used in LQG
\cite{Thiemann:1997rq,Starodubtsev:2002xe,Ashtekar:2002vh}. Note
that there are other diffeomorphism invariant representations
\cite{Kaminski:2005nc,Kaminski:2006ta} that we will not make use of.
We will however use slightly non-standard notation: Denote by
$\avec{\lambda},\avec{\lambda}',\ldots$ functions from $\Sigma$ to
$\ind$ that are non-zero at most in a finite number of points, and
for a given point $x\in \Sigma$, let $\lambda_x,\lambda_x',\ldots$
denote their values at $x$. Let $\cyl$ be the free linear space over
such functions, and write the generators as
$\ket{\avec{\lambda}},\ket{\avec{\lambda}'},\ldots$. An inner
product on $\cyl$ is given by
\begin{equation}
\scpr{\avec{\lambda}}{\avec{\lambda'}}=\prod_{x\in \Sigma}
\delta(\lambda_x,\lambda'_{x}),
\end{equation}
and linear extension, and $\hkin$ is the closure of $\cyl$ under the
associated norm. A representation of the basic quantities is given by
\begin{equation*}
\widehat{T}_{x,\lambda}\ket{\avec{\lambda}}=\ket{\avec{\lambda}+\lambda\avec{\delta}^x},\qquad
\widehat{\pi}(f)\ket{\avec{\lambda}}=\sum_{x\in\Sigma} \lambda_x
f(x)\ket{\avec{\lambda}},
\end{equation*}
where we have denoted with $\avec{\delta}^x$ the Kronecker-delta
$\kron{\cdot}{x}.$
We will also call supp$(\avec{\lambda})$ the \textit{graph
underlying} $\ket{\avec{\lambda}}$.

Diffeomorphisms $\varphi$ act as unitary operators $U_{\varphi}$ on
$\hkin$: Denote with $\varphi\ast\avec{\lambda}$ the pullback
 $\avec{(\lambda\circ\varphi)}$ under diffeomorphisms, then
$U_\varphi\ket{\avec{\lambda}}=\ket{\varphi^{-1}\ast\avec{\lambda}}$.
%------------------------------------------------------------------
\section{The diffeomorphism invariant Hilbert space $\hdiff$}
\label{se_hdiff}
%------------------------------------------------------------------
In this section we will define a Hilbert space of spatially
diffeomorphism invariant states for the scalar field. At first it
might be surprising that there would be something non-trivial left
after implementation of the diffeomorphism constraint\footnote{We
are grateful to T.\ Thiemann for pointing this important issue out
to us.}: After all, heuristic counting would suggest that since the
phase space is coordinatized by two fields, the constraint
hypersurface should be by one, and the reduced phase space by none.
In Appendix \ref{ap_reduced}, we will however give a (heuristic)
argument to the effect that the reduced phase space is in fact
rather large.

We will define the  Hilbert space $\hdiff$ of diffeomorphism
invariant states in analogy with full LQG \cite{Ashtekar:1995zh}.
There it is part of the dual of
$\cyl$, equipped with a scalar product. The latter is defined using
group averaging, which gives a map from $\cyl$ to diffeomorphism
invariant elements of $\cyl^*$. On a heuristic level, that map is
obtained by averaging over all diffeomorphisms,
\begin{equation*}
(\Gamma \Psi)(\Phi) = (\text{Vol}(\diff))^{-1} \int_{\diff}
D\varphi\, \scpr{U_\varphi\Psi}{\Phi}.
\end{equation*}
The actual formula is a bit more subtle. To spell it out let us
introduce some concepts (we follow the exposition in \cite{Ashtekar:2004eh}): Let
$\diff$ be a group of diffeomorphisms (analytic, semi-analytic,
\ldots), and given a graph $\gamma$, denote by $\diff_{\gamma}$ the
subgroup of diffeomorphisms mapping $\gamma$ onto itself. Narrowing
it down even further, let $\tdiff_\gamma$ be the subgroup of $\diff$
which is the identity on $\gamma$. The quotient $\gs_\gamma
\doteq\diff_\gamma / \tdiff_\gamma$ is called the set of
$\textit{graph symmetries}$. With these definitions, the actual
result for a function $\Psi_\gamma$ cylindrical on
$\gamma$ is
\begin{equation}
\label{eq_rig1}
 (\Gamma \Psi_\gamma)(\Phi)=\sum_{\varphi_1\in \diff/\diff_\gamma}\frac{1}{\betr{\gs_\gamma}}
 \sum_{\varphi_2\in \gs_\gamma}\scpr{\varphi_1\ast\varphi_2\ast \Psi_\gamma}{\Phi}.
\end{equation}
The division by the order of $\gs_\gamma$ in \eqref{eq_rig1} is
necessary for consistency in the case of full LQG, but for the scalar
field alone there is no clear justification. Thus, to be most
general, we will consider all rigging maps
\begin{equation}
\label{eq_rig2}
 (\Gamma_F\Psi_\gamma)(\Phi)\doteq\sum_{\varphi_1\in \diff/\diff_\gamma}F(\betr{\gs_\gamma})
 \sum_{\varphi_2\in \gs_\gamma}\scpr{\varphi_1\ast\varphi_2\ast \Psi_\gamma}{\Phi}
\end{equation}
with $F(n)$ a strictly positive function on $\N$. Their images
$\Gamma_F(\cyl)$ are all the same, but they induce different inner
products
\begin{equation}
\label{eq_induced}
\scprd{\Gamma_F\Psi}{\Gamma_F\Psi'}_F\doteq(\Gamma_F\Psi)(\Psi')
\end{equation}
and hence different closures $\hdiff^F$.\footnote{To be precise and
avoid confusion: The $\hdiff^F$ are certainly all equivalent
\textit{as Hilbert spaces}, but they may differ as closures
\textit{of $\cyl^*$}.}

Our next task will be to describe the structure of these Hilbert
spaces much more explicitly in the case of the scalar field. The
results will be contained in Proposition \ref{pr_hdiff} and
Corollary \ref{co_hdiff}. Before getting there, we need some more
preparations: We have to talk about the class of diffeomorphisms that
we admit. Several groups have been considered (piecewise linear,
semi-analytic, analytic, smooth etc.) and we will not describe them
in any detail. Rather we will base the rest of the article on the
following assumption:
\begin{assu}
\label{as_diff} For any two ordered sets $(p_1,\ldots,p_n)$,
$(p'_1,\ldots,p'_n),$ of $n$ points of $M$ there is
$\varphi\in\diff$ such that $\varphi(p_i)=p'_i, i=1,\ldots n$.
\end{assu}
This assumption is certainly wrong for $\dim M=1$. For analytic
diffeomorphisms, it seems hard to prove or disprove it, even in
special cases. On the other hand it is probably true for the smooth
category and $\dim M>1$.\footnote{Let us sketch a proof in the latter
case: Given ordered sets of $n$ points $(p_i),(p'_i)$, we want to find $\varphi$ such that
$\varphi(p_i)=p'_i$.
\begin{enumerate}
\item We can find mutually non-intersecting continuous paths $e_i$
connecting $p_i$ with $p'_i$ (Proof by induction:)
\item Blow up paths $e_i$ to mutually non-intersecting tubes $T_i$.
\item Construct to each $T_i$ a diffeomorphism $\varphi_i$ that is
the identity outside of $T_i$ and maps $p_i$ to $p_i'$. Then
$\varphi=\varphi_1\ast\varphi_2\ast\ldots$ does the job.
\end{enumerate}}

The important consequence of the assumption is the following: The
diffeomorphism invariant information contained in a set of pairs of
points of $\Sigma$ and elements of $\ind$ is reduced to how many
times each element of $\ind$ showed up. This simplifies the
description of the diffeomorphism invariant Hilbert space
drastically.

To make this precise let us introduce some notation: Let
$\ind^*$ denote $\ind\setminus 0$, and
$\mathcal{N}$ the set of functions
$\avec{N}:\ind^*\longrightarrow\N$, zero on all but
finitely many elements of $\ind^*$. Evaluation of such a function
$\avec{N}$ on $\lambda\in \ind^*$ will be denoted by $N_\lambda$.
Consider the free vector space $V_\mathcal{N}$ over the symbols
$\ketd{\avec{N}}$, $\avec{N}\in\mathcal{N}$ and equip it with a
scalar product by stipulating
\begin{equation}
\label{eq_scpr} \scprd{\labe{N}}{\labe{N'}}\doteq \prod_{\lambda\in\ind^*}
\lab{N}{\lambda}!\, \kron{\lab{N}{\lambda}}{\lab{N'}{\lambda}},
\end{equation}
and (sesqui-)linear extension. Completion gives a Hilbert space
$\overline{V}_\mathcal{N}$ with orthogonal basis
$\{\ketd{\labe{N}}\}_{\labe{N}\in \mathcal{N}}$.

We can identify elements of $V_\mathcal{N}$ with diffeomorphism
invariant elements of $\cyl^*$ as follows: For a basis state
$\ket{\avec{\lambda}}$, let
\begin{equation}
\label{eq_ident}
N^{(\avec{\lambda})}_{\lambda'}\doteq \sum_{x\in\Sigma}
\kron{\lambda_x}{\lambda'},
\end{equation}
which is nothing else than saying that
$\lab{N^{(\avec{\lambda})}}{\lambda}$ is is the number of points in
the graph of $\avec{\lambda}$ with label $\lambda'$. Then for
$\brad{\avec{N}}$ set
\begin{equation*}
\brad{\labe{N}}(\ket{\avec{\lambda}})\doteq\scprd{\labe{N}}{\labe{N}^{(\avec{\lambda})}}
\end{equation*}
and extend linearly to all of $\cyl$, and anti-linear to all of
$V_\mathcal{N}$.
\begin{prop}
\label{pr_hdiff} Provided Assumption \ref{as_diff} holds, the
explicit form of the rigging map $\Gamma_{1}$ is
\begin{equation*}
\Gamma_{1}\ket{\avec{\lambda}}=\brad{\avec{N}^{(\avec{\lambda})}}.
\end{equation*}
The inner product \eqref{eq_induced} induced by this map coincides
with the inner product \eqref{eq_scpr}.
\end{prop}
The proof is contained in Appendix \ref{ap_proof2}.

Let us define some operators on $\hdiff^{1}$ by giving their matrix
elements in the basis \eqref{eq_scpr}:
\begin{equation*}
\widehat{N}_\alpha\ketd{\labe{N}}\doteq
\lab{N}{\alpha}\ketd{\labe{N}}\qquad
\widehat{N}\doteq\sum_{\alpha\in\ind^*} \widehat{N}_\alpha\qquad
\widehat{N}!\ketd{\labe{N}}\doteq\big(\sum_{\alpha\in\ind^*}
N_\alpha\big)!\,\ketd{\labe{N}}
\end{equation*}
One can easily check that $\widehat{N}_\alpha$ (and hence
$\widehat{N}$) and $\widehat{N}!$ are symmetric on $V_\mathcal{N}$.
\begin{coro}
\label{co_hdiff} Provided Assumption \ref{as_diff} holds, the general
rigging map is given by
$\Gamma_{F}\ket{\avec{\lambda}}=\brad{\labe{N}^{(\avec{\lambda})}}F(\widehat{N}!)$.
The resulting scalar product is
\begin{equation}
  \scprd{\cdot}{\cdot}_F = \scprd{\cdot}{1/F(\widehat{N}!)\,\cdot}.
\end{equation}
\end{coro}
\begin{proof}
It is clear from the definition \ref{eq_rig2} of $\Gamma_F$ that
$\Gamma_F=F(\wh{N}!)\circ\Gamma_1$. Then for the inner product \eqref{eq_induced}
induced on $V_\mathcal{N}$ we have
\begin{align*}
\scprd{\Gamma_F(\ket{\avec{\lambda}})}{\Gamma_F(\ket{\avec{\lambda'}})}_F
&=(\Gamma_F\ket{\avec{\lambda}})(\avec{\lambda'})\\
\Longleftrightarrow \quad\scprd{F(\wh{N}!)\avec{N}^{(\lambda)}}{F(\wh{N}!)\avec{N}^{(\lambda')}}_F
&=\scprd{F(\wh{N}!)\avec{N}^{(\lambda)})}{\avec{N}^{(\lambda')}}.
\end{align*}
But since the image of $\Gamma_F$ is $\cyl$, and $F(\wh{N}!)$ maps $\cyl$ to itself and is
invertible, we can conclude
\begin{equation*}
\scprd{\Psi}{\Psi'}_F=\scpr{\Psi}{1/F(\wh{N}!)\Psi'}
\end{equation*}
for any $\Psi,\Psi'\in\cyl$.
\end{proof}
To finish our exposition, we define some additional natural operators on
$\hdiff^{F=1}$: For $\alpha\in\ind^*$,
\begin{equation*}
\ann{\alpha}\ketd{\labe{N}}\doteq
N_\alpha\ketd{\labe{N}-\delta_\alpha},\qquad
\cre{\alpha}\ketd{\labe{N}}\doteq\ketd{\labe{N}+\delta_\alpha}.
\end{equation*}
$\ann{\alpha}$, $\cre{\alpha}$ are mutually adjoint. Furthermore one
finds the familiar relations
\begin{align}
[\cre{\alpha},\cre{\alpha'}]&=[\ann{\alpha},\ann{\alpha'}]=0
&[\ann{\alpha},\cre{\alpha'}]&=\kron{\alpha}{\alpha'} \text{id}\\
\cre{\alpha}\ann{\alpha}&=\widehat{N}_\alpha
\end{align}
for $\alpha,\alpha'\in\ind^*$.
Let us conclude with a few remarks.\\
1) The picture that we
obtained is very simple and combinatorial. All that survives group
averaging is multiplicity of labels. This means in particular that on the
diffeomorphism invariant level, the quantum theory for the scalar
field has ``forgotten" about the dimension of the underlying spatial
manifold $\Sigma$ whenever Assumption \ref{as_diff} holds.\\
2) There is a natural way to write $\hdiff^1$ as a Fock space: Let
$h=L^2(\ind^*,\text{d}\mu_{\text{discr}})$, where
$\text{d}\mu_{\text{discr}}$ is the discrete measure on $\ind^*$.
Then there is a natural isomorphism between the eigenspace to the
eigenvalue $n$ of $\widehat{N}$ and the symmetric tensor product
$\otimes^n_S h$, and hence between $\hdiff$ and $\mathcal{F}_S( h
)$. The natural annihilation and creation operators on
$\mathcal{F}_S(h)$, evaluated on elements $\delta(\cdot,\alpha)$ of
$h$ are mapped under this
isomorphism to the $\ann{\alpha},\cre{\alpha}$ defined above.\\
However we
should point out that the interpretation of the $\lambda$ labels is
very different from that of the (Fourier-)$k$ labels encountered in
the standard Fock-quantization of a free scalar field. The
$\lambda\in \ind$ are representation labels for the group that plays
the role of ``target space", whereas the $k$ of the standard
quantization come from Fourier transform \textit{on} the spatial manifold
$\Sigma$.\\
3) We note that $F(\wh{N}!)^{1/2}$ is a unitary map from $\hdiff^1$ to $\hdiff^F$.
Thus the annihilation and creation operators etc. can be carried over from $\hdiff^1$
to any $\hdiff^F$. (However this is somewhat unnatural, as the unitary map
does not come from a unitary map on the Hilbert space $h$.)\\
4) Finally
equation \eqref{eq_scpr} suggests that $\hdiff^1$ can be written as an
infinite direct product of Hilbert spaces. We will not investigate
this further.
%------------------------------------------------------------------
\section{Quantization on $\hdiff$}
\label{se_quant}
%------------------------------------------------------------------
In this section we will discuss the quantization of the quantities
$L_\alpha$ in \eqref{eq_thel}. I.e. we are looking for operators
$\widehat{L}_\alpha$ (on $\hkin$ or $\hdiff^{F}$) such that
\begin{equation}
\label{eq_good} \comm{\widehat{L}_\alpha}{\widehat{L}_{\alpha'}}
=(\alpha'-\alpha)\widehat{L}_{\alpha+\alpha'},\qquad
(\widehat{L}_\alpha)^\dagger=\widehat{L}_{-\alpha}.
\end{equation}
The obvious difficulty of this endeavor is that the $L_\alpha$ depend
on both, configuration and momentum variables, and hence a choice of
ordering will have to be made.

In the present article, we will consider one particular ordering, the
symmetric (or Weyl-) ordering. We will not define this order in any
generality, but instead give an example from quantum mechanics that
nevertheless contains all we need to know here: The
symmetric ordering for a phase space function $f(x)p$ is
\begin{equation}
\label{eq_order}
\widehat{f(x)p}=\frac{1}{2}(\widehat{p}f(\widehat{x})+
f(\widehat{x})\widehat{p})
=\frac{1}{2}(f(\widehat{x})\widehat{p}+[\overline{f}(\widehat{x})\widehat{p}]^\dagger)
=f(\widehat{x})\widehat{p}+\frac{1}{2i}f'(\widehat{x}).
\end{equation}
It is instructive to start with trying to quantize the $L_\alpha$ on
$\hkin$. Consideration of the actions of $\widehat{\pi}(x)$ and
$\wh{T}_{x,\lambda}$ on $\hkin$, and ordering $\pi$ to the right gives an
operator
\begin{equation}
\label{eq_thes}
\widehat{S}_\alpha\ket{\avec{\lambda}}\doteq\sum_{x\in\Sigma}
\lambda_x \ket{\avec{\lambda}+ \alpha\avec{\delta}^x},
\end{equation}
where we have denoted with $\avec{\delta}^x$ the Kronecker-delta
$\kron{\cdot}{x}.$ A check on the commutation relation turns out
positive: Since
\begin{equation*}
\widehat{S}_\alpha\widehat{S}_{\alpha'}\ket{\avec{\lambda}}
=\sum_{x\neq x'} \lambda_x\lambda_{x'}
\ket{\avec{\lambda}+ \alpha\avec{\delta}^x+\alpha'\avec{\delta}^{x'}}
+ \sum_{x}\lambda_x(\lambda_x+\alpha') \ket{\avec{\lambda}+
(\alpha+\alpha')\avec{\delta}^x},
\end{equation*}
we find that
\begin{align*}
\comm{\widehat{S}_\alpha}{\widehat{S}_{\alpha'}}\ket{\avec{\lambda}}
=&\sum_{x}\left(\lambda_x(\lambda_x+\alpha')-\lambda_x(\lambda_x+\alpha)\right)
\ket{\avec{\lambda}+ (\alpha+\alpha')\avec{\delta}^x}\\
=&(\alpha'-\alpha)\widehat{S}_{\alpha+\alpha'}\ket{\avec{\lambda}}.
\end{align*}
So the $\widehat{S}_\alpha$ do fulfill the right commutation
relations. What about the adjointness relation? We expect it not be
fulfilled since we have ordered the $\pi$ to the right. Symmetric
order \eqref{eq_order} would ask for the operator
\begin{equation*}
\wh{L}_\alpha\doteq
\frac{1}{2}(\widehat{S}_\alpha+\widehat{S}^\dagger_{-\alpha})
\end{equation*}
($\wh{S}_\alpha$ corresponds to $f(\wh{x})\wh{p}$ in \eqref{eq_order}, so the
above corresponds to the second equality in \eqref{eq_order}).
So what is $\widehat{S}^\dagger_\alpha$? Alas
\begin{prop}
\label{pr_adjoint}
No element of $\cyl$ is in the domain of definition of
$\widehat{S}^\dagger_\alpha$ for $\alpha\neq 0$ (where we take
dom($\widehat{S}_\alpha$) to be $\cyl$).
\end{prop}
\begin{proof}
Let $\alpha\neq 0$. Given any $\ket{\avec{\lambda}}\in\cyl$, if we show that
there are uncountably infinitely many $\ket{\avec{\lambda}'}$ such that
$\scpr{\avec{\lambda}}{\widehat{S}_\alpha\avec{\lambda}'}\neq 0$ we are done, for
then $\scpr{\avec{\lambda}}{\widehat{S}_\alpha\,\cdot\,}$ can not be
bounded.\newline Indeed, given $\ket{\avec{\lambda}}$, choose $x$ such that
$\lambda_x=0$ and define $\avec{\lambda}'=\avec{\lambda}-\alpha\avec{\delta}^x$. Then
$\scpr{\avec{\lambda}}{\widehat{S}_\alpha\avec{\lambda}'}=1$. But there are
uncountably many $x$ with $\lambda_x= 0$.
\end{proof}
We can certainly guess where this failure comes from: One has to expect
that
\begin{equation*}
\wh{S}_\alpha^\dagger=\wh{S}_{-\alpha}-\alpha\sum_x\wh{T}_{x,-\alpha}
\end{equation*}
if it existed, but it can't because the sum over $x\in\Sigma$
does not converge on $\cyl$.

We will now consider representing the $L_\alpha$ on $\hdiff^F$. We
start by noting that any operator $a$ on $\hkin$ that maps $\cyl$ to
$\cyl$ automatically has a dual action on  $\cyl^*$: For $l\in
\cyl^*$ set $\dual{a} l \doteq l\circ a$. This dual action preserves
commutation relations in the sense that $[\dual{a},
\dual{b}]=\dual{[b,a]}$. Furthermore, if the operator $a$ is
diffeomorphism-invariant then $\dual{a}$ preserves $V_\mathcal{N}$.
Thus in particular for the operators $\wh{S}_\alpha$ \eqref{eq_thes} we
get a dual action $\dual{S}_\alpha$ on $V_\mathcal{N}$. It is
important to stress that at this point there is no need for any
(choice of) rigging map. It will however become important once we
consider adjoints.

What is the explicit form of the $\dual{S}_\alpha$?
%%%%%%%%%%%%%%%%%%%%%%%%%%%%%%%%%%%%%%%%%%%%%%%%%%%%%%%%
\begin{prop}
\label{pr_dualaction} The action of $\dual{S}_\alpha$ on
$V_\mathcal{N}$ can be expressed as
\begin{equation}
\label{eq_dual} \dual{S}_\alpha= -\alpha \cre{-\alpha} +
\sum_{\lambda\neq 0} (\lambda -\alpha)
\cre{\lambda-\alpha}\ann{\lambda}
\end{equation}
\end{prop}
%%%%%%%%%%%%%%%%%%%%%%%%%%%%%%%%%%%%%%%%%%%%%%%%%%%%%%%%
The proof of this proposition is contained in Appendix
\ref{ap_proof}.

It will be convenient in the following to adopt the notation
$\ann{0}=\cre{0}\doteq\text id$, the identity map. Then we can write
more compactly
\begin{equation*}
\dual{S}_\alpha= \sum_{\lambda} (\lambda -\alpha)
\cre{\lambda-\alpha}\ann{\lambda}.
\end{equation*}
Now, in contrast to the situation on $\hkin$, the $\dual{S}_\alpha$
(their domain taken to be $V_{\mathcal{N}}$)
possess adjoints that are nontrivial on $V_{\mathcal{N}}$, under all of the scalar
products (indexed by $F$)
considered in the preceding section. Let us start with $F=1$: One
finds immediately (we remind the reader that where we denote the
adjoint for $F=1$ with $^\dagger$)
\begin{equation*}
\dual{S}_\alpha^\dagger = \sum_{\lambda} \lambda
\cre{\lambda+\alpha}\ann{\lambda}=\dual{S}_{-\alpha}-
\alpha\sum_\lambda\cre{\lambda+\alpha}\ann{\lambda}
\end{equation*}
on $V_{\mathcal{N}}$.\footnote{From here on, we will not explicitly
mention anymore that we consider all adjoint operators restricted to $V_{\mathcal{N}}$.}
Thus the $\dual{S}_\alpha$ do not yet fulfill the right adjointness
relations. But that is no surprise as we have not yet ordered
symmetrically. Let
\begin{equation*}
\widehat{L}_\alpha\doteq\frac{1}{2}\left(\dual{S}_{-\alpha} +
\dual{S}_{\alpha}^\dagger\right)
=\dual{S}_{-\alpha}-\frac{\alpha}{2}\sum_\lambda\cre{\lambda+\alpha}\ann{\lambda}=
\sum_\lambda
\left(\lambda+\frac{\alpha}{2}\right)\cre{\lambda+\alpha}\ann{\lambda}.
\end{equation*}
By
definition we now have
$\widehat{L}_\alpha^\dagger=\widehat{L}_{-\alpha}$. However the
commutation relations have to be checked. A short calculation shows
that
\begin{equation*}
[\widehat{L}_\alpha,\widehat{L}_{\alpha'}]=(\alpha'-\alpha)\widehat{L}_{\alpha+\alpha'}
+\frac{1}{4}\alpha\alpha'\left(
\cre{\alpha}\ann{-\alpha'}-\cre{\alpha'}\ann{-\alpha} \right),
\end{equation*}
i.e.~ an anomalous term appears on the right hand side.

We can follow the same steps, but for the more complicated situation
$F(n)\neq 1$. In this case, the dual action \eqref{eq_dual} of the
$\wh{S}_\alpha$ remains the same, however the scalar product, and hence
the notion of adjoint, and of symmetric ordering, change. Let us
denote the adjoint with respect to $\scprd{\cdot}{\cdot}_F$ for
$F\neq 1$ with $^*$. Then symmetric ordering gives
\begin{equation*}
\wh{L}'_\alpha
\doteq\frac{1}{2}(\widetilde{S}_{-\alpha}+\widetilde{S}_{\alpha}^*).
\end{equation*}
Again, adjointness relations are automatic. For the commutators we
find
\begin{prop}
\label{pr_general}
\begin{equation*}
\comm{\wh{L}'_\alpha}{\wh{L}'_{\alpha'}}=(\alpha'-\alpha)\wh{L}'_{\alpha+\alpha'}+\frac{\alpha\alpha'}{4}
(\cre{\alpha}(\Delta(\widehat{N})+1)\ann{-\alpha'}-\cre{\alpha'}(\Delta(\widehat{N})+1)\ann{-\alpha})
\end{equation*}
where $\Delta(n)=F((n+1)!)/F((n+2)!)-F(n!)/F((n+1)!)$.
\end{prop}
Let us finish with three remarks:\\
1) The standard choice
$F(n)=1/n$ does not lead to a vanishing of the unwanted term, as
in that case $\Delta(n)+1=1+n+2-(n+1)=2\neq 0$.\\
2) One can ask the question if there is a valid (i.e. strictly
positive) choice of $F$ that leads to $\Delta(n)+1=0$. It is not hard
to solve the latter equation: Let $f(n)=F(n!)$. Then  the equation
immediately implies
\begin{equation*}
\frac{f(n)}{f(n+1)}= c_0-n
\end{equation*}
for some $c_0$ whence
\begin{equation*}
f(n)^{-1}= f(0)^{-1}c_0(c_0-1)(c_0-2)\ldots(c_0-(n-1)).
\end{equation*}
One sees immediately that there is no strictly positive solution. The
best one can hope for is a non-negative solution, which is achieved
by choosing for $c_0$ some integer $N_0$,
\begin{equation*}
1/f(n)=\begin{cases}f(0)N_0!/(N_0-n)! & \text{ for $n\leq N_0$
}\\0&\text{ else}\end{cases}.
\end{equation*}
Since $1/f(n)$ can actually become zero, the Hilbert space that one
would obtain would be very different from the $\hdiff$ considered so far. (In particular,
$f(\widehat{N})^{-1/2}$ would not be a unitary map to $\hdiff^{F=1}$.)
Also, it is in fact not immediately clear wether Proposition
\ref{pr_general} still holds in this case. We do however suspect that
this choice would yield a representation of the relations
\eqref{eq_good}. The reason is that there is a natural interpretation
for this inner product: It seems to arise if, instead of $\Sigma$
being a differentiable manifold, one allows $\Sigma$ to be just a
finite set, of $N_0$ points. Diffeomorphisms would be replaced by
permutations, and group averaging can be carried out explicitly,
leading to precisely the above inner product. Note however that
$(N_0-n)!$ is not the number of graph symmetries (that would be $n!$)
but the number of permutations that leave the graph invariant.\\
3) The unwanted term is really unwanted, because it is neither
central, nor can it be expressed in any simple way through the
$\wh{L}_\alpha$. It is thus not a simple ``quantum correction".
%------------------------------------------------------------------
\section{Closing remarks}
\label{se_close}
%------------------------------------------------------------------
In the preceding sections we have, under Assumption \ref{as_diff},
explicitly worked out the structure of
$\hdiff^F$ for a scalar field, both for the standard choice of
combinatorial factor $F$, as well as for a large class of other choices.
Thereby we have seen that the role of the group averaging map is quite subtle:
Different maps give isomorphic Hilbert-spaces, but they are not equal as
spaces of linear functionals over $\cyl$. Thus if the latter is used
for quantization, results in general depend on the map.
We have also shown that there is a natural Fock structure on $\hdiff^1$
(which can, by unitary equivalence, also be defined on any $\hdiff^F$).\\
Do these findings have any implications for the full theory? Certainly not immediately.
For one thing, in the full theory the structure of $\hdiff$ is much more
complicated. Creating a new loop in a given state can happen in many ways,
so that one does not expect to simply get away with standard annihilation
and creation operators. There is knotting, the analog of which in the
situation here would be to work with $\Sigma$ one dimensional, i.e.
without Assumption \ref{as_diff}. Our results are not
even directly applicable to a scalar field coupled to gravity, because
$\hdiff$ in that case would \textit{not} be the tensor product of the diff-invariant
Hilbert space for gravity and that for the scalar field.
But the results can give some inspiration for looking at the diff-invariant Hilbert
space in LQG in new ways, ex. building it up by performing simple
operations on the vacuum etc.

We have furthermore tried to represent the algebra of the $L_\alpha$ on $\hdiff$
(and $\hkin$), using symmetric ordering, without success. There are
several attitudes that one can take towards this failure and its
implications:\newline
1) Symmetric ordering is not right -- maybe a different ordering prescription
can help? Indeed this is very well possible. We have tried to start with an
ordering like
\begin{equation*}
L_\alpha=\int \text{sign}(\pi(x))\sqrt{|\pi|}(x)\exp(i\alpha\phi(x))\sqrt{|\pi|}(x)
\end{equation*}
but obtained more complicated correction terms. But maybe one can come up
with a better prescription.\newline
2) Maybe these quantities simply fail to exist at all as operators, on any of the $\hdiff^F$?
That seems to be plausible, too. After all, at least on the kinematical Hilbert space
one can understand the reason for the failure of the quantization very clearly:
It is the failure of operators like
\begin{equation*}
\int_\Sigma \comm{\widehat{\pi}(x)}{\exp i\alpha\widehat{\phi}(x)}
\end{equation*}
to make sense. And there is no obvious reason why this problem should go away
on the diffeomorphism-invariant level.\newline
3) Maybe one should not care so much about the correction term or about the adjointness relations?
It is certainly hard to argue
about the importance of this term because the whole set-up is very un-physical.
We would however like to repeat, however, that the correction term is not central,
is not expressible in any straightforward way through the $L_{\alpha}$'s,
and therefore generates a whole bunch of new quantum objects that have
no classical analogue. Abandoning the adjointness relations does a very similar
thing in one stroke, namely doubling the number of members of the algebra.

Do our results have implications for physically more interesting
situations? We do not know. One thing that may be noted is that
asking for diffeomorphism invariant operators to be defined on the
kinematical Hilbert space already (ex. \cite{Ashtekar:1995zh}) may
be too much (see Prop. \ref{pr_adjoint}). Furthermore it is
certainly true that there are somewhat similar ordering issues also
for physically very important operators such as the Hamiltonian
constraint. The situation in that case is however also different in
many respects (and it has been argued that violation of
adjointness-relations may be necessary in that case for other
reasons), so that direct conclusions can not be drawn.

When comparing the present paper to the literature on the
implementation of the Hamilton constraint, one should keep in mind
that we asked for more than could be, and has been, asked in that
case: Since the algebra we considered here closed, we could ask for
its implementation. Since the quantities were diffeomorphism
invariant, we could seek the representation on $\hdiff$. Neither
``luxuries" are available for the Hamiltonian constraint. Asking for
a lot makes it harder to succeed, and we saw that in the present
situation it was hard to have both, anomaly freedom and correct
adjointness relations. One can certainly ask for less, whence it
would be easier to declare full success.

In any case we should stress again that the situation for the
implementation of the Hamiltonian constraint is very different from
the one considered here, and moreover there are many suggestions
(ex. \cite{Thiemann:2005zg}) as to how to accomplish the former that
differ on a technical and conceptual level from the present
approach. Therefore direct conclusions from our work here for the
issue of the dynamics of LQG can unfortunately not be drawn.

We do however feel that seeing that things are not very
straightforward even in a simple case such as considered here,
should make one be extremely cautious in more complicated
situations.

The present work could be extended in many ways: Maybe there is another quantization
procedure that leads to more satisfying results?
Can the analogy with quantization techniques used in full LQG be strengthened?\newline
It would also be interesting to study the situation for $\Sigma$ one dimensional.
In this case, $\hdiff$ is more complicated, and thus the previous analysis does not apply.
In this case, there may even be some physical application, as one could use the likes
of the $L_\alpha$ to build operators that resemble simple vertex operators for the
bosonic string.

%---------------------------------------------------------
\section*{Acknowledgements}
We would like to thank members of the Institute for Gravitational Physics
at Penn State University
for hospitality and discussions, and also Thomas Thiemann  for discussions and
valuable input.

%---------------------------------------------------------
\begin{appendix}
%---------------------------------------------------------
\section{A look at the reduced phase space}
\label{ap_reduced} The constraint is $\pi(x)\partial_a\phi(x)=0$. So
for $(\pi,\phi)$ on the constraint surface $\pi(x)=0$ or
$\phi(x)=const.$ at any given point $x$. Identifying points on the
constraint surface that can be mapped onto each other by the maps
generated by the constraint gives the reduced phase space. How large
is it?\footnote{We are grateful to T. Thiemann for posing this
question.} We will not answer this question directly, but at least
for the smooth category we will exhibit enough points such that all
the functions $L_\alpha$ are necessary to tell them apart: Given a
sequence $s=(v_i,\phi_i)$ of pairs of real numbers we construct a
point on the constraint surface as follows: Choose non-intersecting
open subsets $U_i$ of $\Sigma$. Then choose a phase space point
$(\phi_s,\pi_s)$ such that
\begin{equation*}
\pi_s\rvert_{\Sigma-\cup_i U_i }=0,  \qquad\int_{U_i}\pi = v_i,\qquad
\phi\rvert_{U_i}=\phi_i
\end{equation*}
Such $(\phi_s,\pi_s)$ obviously satisfies the constraint. Thus to
each sequence $s$ we find at least one point on the constraint
surface and thus a point in the reduced phase space. Are these points
distinct? We compute the quantities $L_\alpha$ on $(\phi_s,\pi_s)$
and find
\begin{equation*}
L_\alpha(\phi_s,\pi_s)=\sum_j v_j\exp(i c_j \alpha)=:F_s(\alpha)
\end{equation*}
Let us assume $s$ is chosen such that the above converges uniformly
in $\alpha$. Since the $L_\alpha$ are invariant under
diffeomorphisms, there are at least as many points in the reduced
phase as there are functions $F_s$ of the above form, that is
\begin{itemize}
\item for $\ind=\Z$: At least as many points as functions on a circle with
uniformly converging Fourier series.
\item for $\ind=\R$: At least as many points as almost periodic functions.
\end{itemize}
It should also be clear that one indeed needs to consider all
$L_\alpha$ to separate the points thus obtained. (Certainly there are
more points in the reduced phase space, and the $L_\alpha$ are thus
not separating all points.)
%---------------------------------------------------------------------
\section{Proof of Prop. \ref{pr_hdiff}}
\label{ap_proof2}
%--------------------------------------------------------------------
Let us start by writing \eqref{eq_rig2} in the case of $F=1$ for two
basis elements $\ket{\avec{\lambda}}$, $\ket{\avec{\lambda'}}$:
\footnote{Since we will consider the case $F=1$ exclusively, we will
here and in the following drop the corresponding subscript.}
\begin{equation}
\label{eq_scpr1}
\brad{\Gamma(\avec{\lambda})}(\ket{\avec{\lambda}'})=
 \sum_{[\varphi_1]\in \diff/\diff_\gamma}
 \sum_{[\varphi_2]\in \gs_\gamma}\scpr{\varphi_1\ast\varphi_2\ast \avec{\lambda}}{\avec{\lambda}'}.
\end{equation}
The kinematical scalar product in this formula will only be nonzero
if $(\varphi_1\ast\varphi_2\ast \lambda)_x=\lambda'_x$ for all
points $x$. Thus it is important to understand when this can happen.
Let us rewrite
\begin{equation*}
\lambda_{x}= \sum_i \lambda_i\delta(x_i,x),
\end{equation*}
with $\lambda_i\neq 0$ and $x_i\neq x_j$ for $i\neq j$. Let us also
introduce the shorthand $\gamma$ for the set of the $x_i$. We
similarly decompose $\lambda'$, and consider the second sum in
\eqref{eq_scpr1}: The action of elements of $\diff_\gamma$, when
restricted to $\gamma$, can only consist in permutations of the
points of $\gamma$, and per our Assumption \ref{as_diff} all the
permutations do occur. Given two diffeomorphisms
$\varphi_1,\varphi_2\in \diff_\gamma$ that reduce to the same
permutation, then they are in the same equivalence class in
$\gs_\alpha$, as $ \varphi_1\circ\varphi_2^{-1}$ is in $\tdiff$, and
$(\varphi_1\circ\varphi_2^{-1})\circ\varphi_2=\varphi_1$. Hence
$\gs_\gamma$ is isomorphic to the permutation group
$\mathcal{P}_{\betr{\gamma}}$, and in terms of the latter, the action
on states is the obvious:
\begin{equation*}
\ket{\avec\lambda}\longmapsto\ket{\pi\cdot\avec\lambda},\quad\text{
with }\quad \pi\cdot\lambda_{x}=\sum_i \lambda_i\delta(x_{\pi(i)},x)
\end{equation*}
So the second sum in \eqref{eq_scpr1} will essentially be over
permutations in $\mathcal{P}_{\betr{\gamma}}$.

Now consider the sum over $\diff /\diff_\gamma$ in \eqref{eq_scpr1}:
The kinematical scalar product will necessarily be zero if
$\varphi_1$ does not map $\gamma$ to $\gamma'$. If
$\betr{\gamma}\neq\betr{\gamma'}$ then this can never happen, and
\eqref{eq_scpr1} gives zero. On the other hand, if
$\betr{\gamma}=\betr{\gamma'}$, then per Assumption \ref{as_diff}
there exists at least one diffeomorphism mapping $\gamma$ to
$\gamma'$. Let $\varphi_1$ be such a diffeomorphism. How how big is
$[\varphi_1]$? Given $\varphi_1'$ that also maps $\gamma$ to
$\gamma'$, then $(\varphi'_1)^{-1}\circ\varphi_1$ is in
$\diff_\gamma$ and thus $\varphi'_1$ is already contained in
$[\varphi_1]$. Thus there is at most one element in the first sum of
\eqref{eq_scpr1} that gives a nonzero contribution.

Altogether we find that \eqref{eq_scpr1} can be simplified to
\begin{equation}
\label{eq_scalar2}
\brad{\Gamma(\avec{\lambda})}(\ket{\avec{\lambda}'})=
 \sum_{\pi\in \mathcal{P}_\gamma} \begin{cases}0 & \text{if} \betr{\gamma}\neq\betr{\gamma'}\\
 \prod_i\delta(\lambda_{\pi(i)},\lambda'_i) & \text{otherwise}
\end{cases}
\end{equation}
\eqref{eq_scalar2} is nonzero only if the sequences $(\lambda_i)$,
$(\lambda'_j)$ are equal modulo permutations. Expressed more
concisely, \eqref{eq_scalar2} is nonzero only if
${N}^{(\avec{\lambda})}_\mu$ and $N^{(\avec{\lambda}')}_\mu$ are
equal for all $\mu\in\ind$ (recall definition \eqref{eq_ident}). In
that case, several terms in the sum
may give nonzero contribution: The order of the stabilizer subgroup
of permutations, for a set of objects with multiplicity $n_1,
n_2,n_3,\ldots$, is $n_1!n_2!n_3!\ldots$. In our case this means that
the sum actually gives $\prod_\mu N^{(\avec{\lambda})}_\mu!$. Thus we
can rewrite \eqref{eq_scalar2} as
\begin{equation*}
\brad{\Gamma(\avec{\lambda})}(\ket{\avec{\lambda}'})=\prod_{\mu\in\ind^*}
N^{(\avec{\lambda})}_\mu! \delta(N^{(\avec{\lambda})}_\mu,
N^{(\avec{\lambda}')}_\mu).
\end{equation*}
But the right hand side of the above equation is just
$\brad{\labe{N}^{(\avec{\lambda})}}(\avec{\lambda}')$. Since the
equation holds for all basis elements, we have shown that indeed
$\brad{\labe{N}^{(\avec{\lambda})}}=\brad{\Gamma(\avec{\lambda})}$.
%--------------------------------------------------------
\section{Proof of Prop. \ref{pr_dualaction}}
\label{ap_proof}
%---------------------------------------------------------
We want to show that the dual action of the operator $S_\alpha$ is
given by \eqref{eq_dual}. Since we posses a basis for $V_\mathcal{N}$
and an orthonormal basis for $\cyl$, it suffices to verify on elements of these
bases. More precisely, we would like to show that
\begin{equation}
\label{eq_a1}
% \scprd{\dual{S}_\alpha \avec{N}}{\Gamma(\ket{\avec{\lambda}})} \equiv
\brad{\dual{S}_\alpha
\avec{N}}(\ket{\avec{\lambda}}) =\brad{\avec{N}}(
\wh{S}_\alpha\ket{\avec{\lambda}})
%\equiv
%\scprd{\avec{N}}{\Gamma(S_\alpha\ket{\avec{\lambda}})}
\end{equation}
for all $\avec{N},\avec{\lambda},\alpha$, and $\dual{S}_\alpha$ \textit{defined
by} \eqref{eq_dual}. We will prove this the un-imaginative way, i.e.
by explicitly computing both sides and verifying that they indeed are
equal in all cases.

Fix $\alpha$, $\avec{\lambda}$, and $\avec{N}$. For now, let us also
assume $\alpha\neq 0$, and come back to $\alpha=0$ later. We write
$\avec{\lambda}$ in the following way:
\begin{equation*}
\lambda_{x}= \sum_i^{n_0} -\alpha \delta(x_i^{(0)}, x)
+\sum_j\sum_i^{n_j} \lambda_j\delta(x^{(j)}_i,x).
\end{equation*}
To make this decomposition unique, the data has to satisfy some
obvious properties. To be clear, let us spell them out: We want
$\lambda_i\neq\lambda_j$ for $i\neq j$, $\lambda_i\neq -\alpha, \neq
0$, all the $n_j>0$, and all the $x_i^{(j)}$ different. In this
notation
\begin{equation*}
\wh{S}_\alpha \ket{\avec{\lambda}} = \sum_{i}^{n_0}-\alpha
\ket{\avec{\lambda} +\alpha\avec{\delta}^{x_i^{(0)}}} +\sum_j\sum_i^{n_j}
\lambda_j\ket{\avec{\lambda} +\alpha \avec{\delta}^{x_i^{(j)}}}.
\end{equation*}
Also
\begin{equation*}
\ketd{\avec{N}^{(\lambda)}}=\ketd{n_0\avec{\delta}^{-\alpha}+\sum_j
n_j \avec{\delta}^{\lambda_j}}.
\end{equation*}
%and
%\begin{equation*}
%\begin{split}
%\Gamma (S_\alpha \ket{\lambda})= &-\alpha n_0
%\ketd{(n_0-1)\avec{\delta}^{-\alpha}
%+\sum_j n_j \avec{\delta}^{\lambda_j}}\\
%&+\sum_i n_i \lambda_i \ketd{n_0\avec{\delta}^{-\alpha}+\sum_j n_j
%\avec{\delta}^{\lambda_j} +\avec{\delta}^{\lambda_i+\alpha}-\avec{\delta}^{\lambda_i}}.
%\end{split}
%\end{equation*}
Finally with the expression \eqref{eq_dual} for $\dual{S}_\alpha$,
\begin{equation*}
\dual{S}_\alpha\ketd{\avec{N}}=-\alpha\ketd{\avec{N}+\avec{\delta}^{-\alpha}}
+\sum_{\mu\neq 0}
N_{\mu}(\mu-\alpha)\ketd{\avec{N}+\avec{\delta}^{\mu-\alpha}-\avec{\delta}^{\mu}}
\end{equation*}
Thus we can write the right hand side of \eqref{eq_a1}
\begin{equation}
\label{eq_rhs}
\begin{split}
\scprd{\avec{N}}{\Gamma(\wh{S}_\alpha\ket{\avec{\lambda}})}=&-\alpha
n_0\scprd{\avec{N}}{(n_0-1)\avec{\delta}^{-\alpha}
+\sum_j n_j \avec{\delta}^{\lambda_j}}\\
&+ \sum_i \lambda_in_i \scprd{\avec{N}}{n_0\avec{\delta}^{-\alpha}+\sum_j
n_j \avec{\delta}^{\lambda_j} +\avec{\delta}^{\lambda_i+\alpha}-\avec{\delta}^{\lambda_i}}
\end{split}
\end{equation}
and for the left hand side we find
\begin{equation}
\label{eq_lhs}
\begin{split}
\scprd{\dual{S}_\alpha \avec{N}}{\Gamma (\ket{\avec{\lambda}})}
=&-\alpha\scprd{\avec{N}+\avec{\delta}^{-\alpha}}{n_0\avec{\delta}^{-\alpha}+\sum_j n_j \avec{\delta}^{\lambda_j}}\\
&+\sum_{\mu\neq 0}N_\mu(\mu-\alpha)
\scprd{N+\avec{\delta}^{\mu-\alpha}-\avec{\delta}^\mu}{n_0\avec{\delta}^{-\alpha}+\sum_j
n_j \avec{\delta}^{\lambda_j}}
\end{split}
\end{equation}
We now have to show that these two expressions in fact evaluate to
the same number. We do this term by term, and start with the first
term in both expressions, i.e. the one with pre-factor $-\alpha$. The
term in \eqref{eq_rhs} can be written as
\begin{equation*}
-\alpha n_0 N_{-\alpha}!\delta(N_{-\alpha}, n_0-1)\cdot\prod
\text{rest}.
\end{equation*}
The term in \eqref{eq_lhs} gives
\begin{align*}
-\alpha(N_{-\alpha}+1)!\delta(N_{-\alpha}+1,n_0)\cdot&\prod\text{rest}\\
&= -\alpha(N_{-\alpha}+1)N_{-\alpha}!\delta(N_{-\alpha},n_0-1)\cdot\prod\text{rest}\\
&= -\alpha n_0\delta(N_{-\alpha},n_0-1)\cdot\prod\text{rest},
\end{align*}
where we have used $\delta(x,y+a)=\delta(x-a,y)$ and the fact that
$n_i$ can be replaced with $N_{\lambda_i}+1$ thanks to the
corresponding Kronecker-delta. Thus it is equal to the term in
\eqref{eq_rhs}, provided the "rest" is the same. That however should
be clear upon a quick inspection.

Now we proceed to comparing the rest of the terms in \eqref{eq_rhs}
and \eqref{eq_lhs}. What we have to do is: Show that to each term in
\eqref{eq_rhs} there is an equal term in \eqref{eq_lhs}, and the rest
of the terms in \eqref{eq_lhs} vanish. To do this, let us first
exclude certain "non-generic" cases, and consider them separately,
for reasons of notation: Fix $i$, and assume that $\lambda_i
+\alpha\neq\lambda_j$ for all $j$,  and also
$\lambda_i+\alpha\neq-\alpha$. Then we can partially expand the $i$th
term in the sum in \eqref{eq_rhs} as
\begin{equation*}
\begin{split}
\lambda_i n_i \cdot N_{-\alpha}!\,\delta(N_{-\alpha},n_0) \cdot
N_{\lambda_i}!\,\delta(N_{\lambda_i},n_i-1)
&\cdot N_{\lambda_i+\alpha}!\, \delta(N_{\lambda_i+\alpha},1)\cdot\\
&\cdot \prod_{j\neq i}N_{\lambda_j}!\,\delta(N_{\lambda_j},n_j) \cdot
\prod \text{rest}
\end{split}
\end{equation*}
Compare this to the $\mu=\lambda_i+\alpha$ term of \eqref{eq_lhs},
which expands as
\begin{equation*}
\begin{split}
\lambda_i N_{\lambda_i+\alpha} \cdot &
N_{-\alpha}!\,\delta(N_{-\alpha},n_0)
\cdot (N_{\lambda_i}+1)!\,\delta (N_{\lambda_i}+1,n_i)\\
&\cdot (N_{\lambda_i+\alpha}-1)!\delta(N_{\lambda_i+\alpha}-1,0)
\cdot\prod_{j\neq i}N_{\lambda_j}!\,\delta(N_{\lambda_j},n_j) \cdot
\prod \text{rest}.
\end{split}
\end{equation*}
That the latter expression is equal to the former can again be easily
seen by the manipulations used before, and the fact that the "rest"
is in both cases just a product of all $\delta(N_\nu,0)$ for
$\nu\neq-\alpha, \lambda_i,\lambda_i+\alpha,\lambda_j$.

Now we have to show that the rest of the terms in \eqref{eq_lhs} are
vanishing.
%
% For $\mu=\alpha$ the term vanishes due to the pre-factor $(\mu -\alpha)$.
%
Expand the $\mu$-term in $\eqref{eq_lhs}$ as
\begin{equation*}
N_\mu(\mu-\alpha)\cdot
\delta(N_{\mu-\alpha}+1,n_0\delta(-\alpha,\mu-\alpha)+\sum_j n_j
\delta(\lambda_j,\mu-\alpha))\cdot\prod\text{rest}
\end{equation*}
Now, $n_0\delta(-\alpha,\mu-\alpha)+\sum_j n_j
\delta(\lambda_j,\mu-\alpha)$ is zero as long as $\mu\neq 0,
\lambda_i+\alpha$ for any $i$. There is actually no $\mu=0$-term in
\eqref{eq_lhs}, and the $\mu=\lambda_i+\alpha$ terms have already
been dealt with. Thus for the remaining $\mu$, the term contains the
factor $\delta(N_{\mu-\alpha}+1,0)$. Since $N_{\mu-\alpha}\geq 0$,
that factor is identically zero and the $\mu$ term in the sum
vanishes.

Now we turn to the non-generic cases:\\
1) $\lambda_i+\alpha=-\alpha$: Let us examine the $i$-term in the sum
\eqref{eq_rhs}. It can be expanded as
\begin{equation*}
n_i\lambda_i N_{-\alpha}!\kron{N_{-\alpha}}{n_0+1}\cdot
N_{\lambda_i}!\kron{N_{\lambda_i}}{n_i-1}\cdot \text{rest}.
\end{equation*}
Similarly consider the $\mu=\lambda_i+\alpha$ term in \eqref{eq_lhs}.
It expands as
\begin{equation*}
N_{-\alpha}\lambda_i (N_{-\alpha}-1)!\kron{N_{-\alpha}-1}{n_0}\cdot
(N_{\lambda_i}+1)!\kron{N_{\lambda_i}+1}{n_i}\cdot \text{rest},
\end{equation*}
thus the $i$-term of \eqref{eq_rhs} equals the the
$\mu=\lambda_i+\alpha$ term in \eqref{eq_lhs}. Comparison of the
other terms proceeds as in the generic case.\newline 2)
$\lambda_i+\alpha=\lambda_{j_0}$: This case can be treated along
similar lines as the case above: Expand the product for the $i$-term
in \eqref{eq_rhs} and compare with the $\mu=\lambda_i+\alpha$ term in
\eqref{eq_rhs}. The rest of the terms coincide as shown in the
non-generic case. We will refrain from giving the details.

Finally we have to treat the case $\alpha=0$: Given $\avec{\lambda}$
we can write (using the same conventions as above)
\begin{equation*}
\lambda_{x}= \sum_j\sum_i^{n_j} \lambda_j\delta(x^{(j)}_i,x).
\end{equation*}
We also have
$\wh{S}_0\ket{\avec{\lambda}}=\sum_i{n_i\lambda_i}\ket{\lambda}$ and hence
\begin{equation*}
\brad{\avec{N}}(\wh{S}_0\ket{\avec{\lambda}})=\sum_i \lambda_i n_i
\scprd{\avec{N}}{\avec{N}^{\avec{\lambda}}},\quad\text{ and }\quad
\ketd{\avec{N}^{\avec{\lambda}}}=\ketd{\sum_i n_i \avec{\delta}^{\lambda_i}}.
\end{equation*}
We want to show that $\dual{S}_0=\sum_{\mu\neq 0}\mu\cre{\mu}\ann{\mu}$,
and the action of the latter is
\begin{equation*}
\sum_{\mu\neq 0}\mu\cre{\mu}\ann{\mu} \ketd{\avec{N}}=\sum_{\mu\neq 0}
\mu N_\mu \ketd{\avec{N}}.
\end{equation*}
Thus we have to show that
\begin{equation*}
\sum_{\mu\neq 0} \mu N_\mu \scprd{\avec{N}}{\sum_i n_i
\avec{\delta}^{\lambda_i}}=\sum_i \lambda_i n_i \scprd{\avec{N}}{\sum_i n_i
\avec{\delta}^{\lambda_i}}.
\end{equation*}
Both sides are zero if $\avec{N}\neq\sum_i n_i \avec{\delta}^{\lambda_i}$.
If instead equality holds in the last equation, then obviously
$\sum_{\mu\neq 0} \mu N_\mu=\sum_i \lambda_i n_i$, and hence we have
proven the assertion.
%--------------------------------------------------------
\section{Proof of Prop. \ref{pr_general}}
\label{ap_proof3}
%---------------------------------------------------------
Let $f(n)=1/F(n!)$. As in the main text, we will denote the adjoint
with respect to $\scprd{\cdot}{\cdot}_F$ for $F\neq 1$ with $^*$, the
one for $F=1$ with $^\dagger$. For any operator $A$ that maps $V_{\mathcal{N}}$ to $V_{\mathcal{N}}$.
they are related by
\begin{equation*}
A^*=f(\widehat{N})^{-1}A^\dagger f(\widehat{N}).
\end{equation*}
In particular they coincide on such operators that commute with
$\widehat{N}$. For $\cre{\alpha}$ we find
\begin{equation*}
(\cre{\alpha})^*=\frac{f(\widehat{N}+1)}{f(\widehat{N})}\ann{\alpha}
\end{equation*}
Therefor
\begin{equation*}
\widetilde{S}_\alpha^{*}=\widetilde{S}_{\alpha}^\dagger-\alpha(\frac{f(\widehat{N}+1)}{f(\widehat{N})}-1)\ann{-\alpha}
\end{equation*}
and
\begin{equation*}
\wh{L}'_\alpha =
\wh{L}_{\alpha}+\frac{\alpha}{2}(\frac{f(\widehat{N}+1)}{f(\widehat{N})}-1)\ann{\alpha}=:\wh{L}_{\alpha}+\wh{\xi}_\alpha
\end{equation*}
For the commutators containing $\wh{\xi}_\alpha$ we find
\begin{align*}
\comm{\wh{L}_\alpha}{\wh{\xi}_{\alpha'}}&=\frac{\alpha\alpha'}{4}\cre{-\alpha}\Delta(\widehat{N})\ann{\alpha'}
+\frac{\alpha\alpha'}{2}\Delta(\widehat{N})\ann{\alpha}\ann{\alpha'}
-\frac{\alpha}{2}(\frac{\alpha}{2}+\alpha')M(\widehat{N})\ann{\alpha+\alpha'}\\
  \comm{\wh{\xi}_\alpha}{\wh{\xi}_{\alpha'}} &=0
\end{align*}
with $M(n)=f(n+1)/f(n)-1$ and $\Delta(n)=M(n+1)-M(n)$. Thus finally
\begin{align*}
  \comm{\wh{L}'_\alpha}{\wh{L}'_{\alpha'}}
  &=\comm{\wh{L}_\alpha}{\wh{L}_{\alpha'}}+\comm{\wh{L}_{\alpha}}{\wh{\xi}_{\alpha'}}+\comm{\wh{\xi}_{\alpha}}{\wh{L}_{\alpha'}}\\
  &=(\alpha-\alpha')\wh{L}_{\alpha+\alpha'}+\frac{\alpha\alpha'}{4}(\cre{-\alpha}\ann{\alpha'}-\cre{-\alpha'}\ann{\alpha})\\
&\qquad\qquad+\frac{\alpha\alpha'}{4}
(\cre{-\alpha}\Delta(\widehat{N})\ann{\alpha'}-\cre{-\alpha'}\Delta(\widehat{N})\ann{\alpha})
+(\alpha-\alpha')\wh{\xi}_{\alpha+\alpha'}\\
&=(\alpha-\alpha')\wh{L}'_{\alpha+\alpha'}+\frac{\alpha\alpha'}{4}
(\cre{-\alpha}(\Delta(\widehat{N})+1)\ann{\alpha'}-\cre{-\alpha'}(\Delta(\widehat{N})+1)\ann{\alpha}).
\end{align*}
\end{appendix}

%***************************************************
%***************************************************

\end{document}